\pdfoutput=1
\documentclass[]{raa}            

\usepackage{graphicx,times}             
\usepackage{natbib}
\usepackage{amssymb,amsmath}
\bibpunct{(}{)}{;}{a}{}{,}

\usepackage[pagebackref=true]{hyperref}

\begin{document}

  \title{Model - Independent Probe of Cosmic Distance Duality Relation}

   \volnopage{Vol.0 (20xx) No.0, 000--000}      
   \setcounter{page}{1}          

   \author{Savita Gahlaut
   \inst{}
   }

   \institute{Deen Dayal Upadhayaya College, University of Delhi, Sector-3, Dwarka, New Delhi 110078, India \\
 E-Mail: savitagahlaut@ddu.du.ac.in  } 
      
\vs\no
   {\small Received 20xx month day; accepted 20xx month day}

\abstract{  In this paper, cosmic distance duality relation is probed without considering any background cosmological model. The only \textit{a priori} assumption is that the Universe is  described by the Friedmann-Lema$\hat{i}$tre-Robertson-Walker (FLRW) metric 
The  strong gravitational lensing (SGL) data is used to construct the dimensionless co-moving distance function $d(z)$ and latest type Ia supernovae (SNe Ia) Pantheon+ data is used to estimate luminosity distances at the corresponding redshifts $z$. Using the distance sum rule along null geodesics of the FLRW metric, the CDDR violation is  probed in both flat and non-flat  space-time  by considering two  parametrizations for $\eta(z)$, the function generally used to probe the possible deviations from CDDR. The results show that, CDDR is compatible with the observations at a very high level of confidence for linear parametrization in flat Universe. In non-flat Universe too, CDDR is valid within $1\sigma$ confidence interval with a mild dependence of $\eta$ on the curvature density parameter $\Omega_{K}$ . The results for non-linear parametrization also show no significant deviation from CDDR. 
\keywords{Cosmology: Observations --- Cosmology, Gravitational Lensing: Strong --- Physical Data and Processes, Cosmology: Distance Scales}
}

   \authorrunning{S. Gahlaut }            
   \titlerunning{Model-Independent Probe of CDDR }  

   \maketitle

%
%
\section{Introduction}           
\label{sect:intro}

Cosmic distance duality relation (CDDR) is one of the fundamental relations in cosmology based on 
Etherington reciprocity relation (\citealt{Etherington+1933}).
If one assumes the space-time geometry to be Riemannian and the number of photons to be conserved  along the null geodesics, the cosmic distance duality relation (CDDR) relates the luminosity distance ($D_{L}$) and angular diameter distance  ($D_{A}$) at the same redshift $z$ as:
\begin{equation}
\eta(z) \equiv \frac{D_{L}(z)}{D_{A}(z)} (1+z)^{-2} = 1
\end{equation}

The CDDR is valid irrespective of which cosmological model is used to describe the Universe.
Many cosmological observations, such as Cosmic Microwave Background Radiation(CMBR), galaxy clusters and gravitational lensing, are based on this fundamental relation. Further, it has been used, assuming it to be true, to study the galaxy cluster physics i.e. the possible  morphologies of galaxy clusters (\citealt{Hol11,Hol12}), the  temperature profile and the gas mass density profile of galaxy clusters (\citealt{Cao11,Cao16}) and so on. 
However CDDR may be violated due to photon absorption by dust (\citealt{Corasaniti}), coupling of photons with non-standard particles (\citealt{Basset}), gravitational lensing, variations in fundamental constants (\citealt{Ellis+etal+2013}) or simply due to systematic observational errors (\citealt{Holanda}). 
Any violation of it i.e. $\eta(z)\neq 1$, may indicate the presence of new physics in the theory of gravity or the existence of some unaccounted systematic errors in the observations. Thus it is crucial to investigate the validity of CDDR with the latest cosmological observations.

 The validity of CDDR  is investigated in several papers using different cosmological data sets available
 (\citealt{Li+etal+2011}; \citealt{Li+etal+2013}; \citealt{Liao+etal+2016};  \citealt{Ellis+etal+2013}; \citealt{Holanda+Busti+2014}; \citealt{Rana+etal+2017}; \citealt{Ruan+etal+2018}; \citealt{Lyu+etal+2020}).
 Some authors consider a background cosmological model while others take a model-independent approach to test CDDR. A pre-assumption of background cosmological model necessitates the optimization of 'nuisance' parameters in the distance-redshift relation. This requirement can introduce  additional uncertainty and bias which can lead to  conflicting results and/or wider confidence intervals \citep{Melia18}. 
Generally, luminosity distance  is estimated from Type Ia Supernovae (SNe Ia) observations while the angular diameter distance is inferred from baryon  acoustic oscillation data (BAO), Hubble parameter data ($H(z)$), strong gravitational lensing data (SGL), gamma-ray bursts and other astrophysical probes . The point to be noted here is that, to study the validity of CDDR, the angular diameter distance and luminosity distance  need to be measured at the same redshift. To check the consistency of different independent datasets in a non-parametric way , numerical techniques, like Genetic Algorithm, Non-parametric smoothing technique (NPS), LOESS $\&$ SIMEX, Gaussian Processes etc., have been used in literature to study the validity of CDDR (\citealt{Ruan+etal+2018, Nesseris+Bellido+2012, Shafieloo+etal+2013, Rana+etal+2016}).  No significant violation of CDDR compared with the uncertainties of the observations has been reported in these studies. Still continuing to test the CDDR with latest cosmological observations is  important along with looking for systematic errors among the previous analysis.

 SNe Ia are considered to be the standard candles in the Universe, as they are as bright  as a galaxy at their peak, and are often used as distance indicators. 
Recently,  \citet{Scolnic}  published the largest compilation of spectroscopically confirmed type Ia Supernovae, named the Pantheon+ sample. The sample is  from the analysis of 1701 light curves compiled across 18 different surveys of 1550 distinct type Ia supernovae. This sample improves upon the previous samples by not only increasing the sample size and redshift span but also by improving the treatments of systematic uncertainties. 
Further, with new powerful space and ground-based telescopes for imaging and spectroscopic
observations, many new strong gravitational lensing (SGL) systems
have been discovered. SGL systems can be a valuable tool to constrain cosmological parameters if one has a good knowledge of lens mass model. From various surveys (SLACS, S4TM, BELLS and BELLS GALLERY), \citet{Chen+etal+2019}  compiled 161 galaxy-scale strong lensing systems, all early-type galaxies with E or S0 morphologies. In the sample only those lens galaxies were selected which do not have any significant substructure or close massive companion. These two conditions ensured that the lens galaxies are spherically symmetric.   

Taking advantage of the significant improvements in strong gravitational lensing (SGL) and SNe Ia observations, I present a new cosmological model-independent approach to probe the CDDR by jointly considering  the strong gravitational lensing data  and latest Pantheon+ SNe Ia data set. The large data size makes the datasets suitable for the  statistical analysis and the constraints on CDDR are improved. In most works so far, CDDR is tested assuming a flat space time, whereas in this work  CDDR probe is extended to the non-flat spacetime too. Using the SGL data,
the dimensionless co-moving distance function $d(z)$ is constructed  to avoid the bias brought in by redshift incoincidence between the observational data sets in analysis. The continuous function $d(z)$ provides the estimate of luminosity distance $D_{L}(z)$. Comparing the luminosity distances derived from SNe Ia data with that estimated from SGL data, one can investigate the validity of CDDR. I take the parametric approach to probe the validity of CDDR. Using the distance sum rule, I investigate CDDR  in both flat and non-flat spacetime.

The outline of the paper is as follows. The methodology to construct model-independent dimensionless co-moving distance function $d(z)$ using SGL data and combining it with SNe Ia data to estimate $\eta(z)$ is discussed in section 2. In Section 3, SGL data and Pantheon+ data used in this work are presented. Results and conclusions are presented in section 4 and discussion in section 5.

\section{Theory}
In a homogeneous and isotropic Universe, the geometry of the spacetime can be described by the Friedmann-Lema$\hat{i}$tre-Robertson-Walker (FLRW) metric:
\begin{equation}
ds^{2} = c^{2}dt^{2}-a(t)^{2}[\frac{dr^{2}}{1-K r^{2}} + r^{2}(d\theta^{2}+ \sin^{2}{\theta} d\phi^{2})]
\end{equation}
 where $c$ is the speed of light, $a(t)$ is the cosmic scale factor and $K$ is the spatial curvature constant and is $\pm 1$ or  $0$ for a suitable choice of units for $r$. 
 
 The dimensionless transverse co-moving distance $d(z)$ in FLRW cosmology is:
\begin{equation}
d(z) = \left\lbrace\begin{array}{cc} D_{c},&K=0\\ &\\
\frac{1}{\sqrt{|\Omega_{K}}|}\sinh{(\sqrt{|\Omega_{K}|}D_{c})},&K=-1\\ &\\ \frac{1}{\sqrt{|\Omega_{K}|}}\sin{(\sqrt{|\Omega_{K}|} D_{c})},&K=1

\end{array}\right.
\end{equation}
where $\Omega_K = -K c^{2}/{a^{2}_{0}H^{2}_{0}}$ , $D_{c} = \int_{0}^{z}\frac{H_{0}}{H(z')}dz'$, $H(z)$ is Hubble parameter and $H_{0}$ and $a_{0}$ are the present values of Hubble parameter and scale factor respectively. The angular diameter distance is 
\begin{equation}
D_{A}(z) = \frac{c}{H_{0}(1+z)} d(z)
\end{equation}
and the luminosity distance is 
\begin{equation}
D_{L}(z) = (1+z) \frac{c}{H_{0}} d(z)
\end{equation}

\subsection{Strong Gravitational Lensing} Multiple images of a background galaxy (source) appear due to the lensing effect of a galaxy or cluster of galaxies (lens) along the line of sight. A ring like structure, called Einstein ring, is formed if the source, lens and observer are perfectly aligned along the same line of sight.
The multiple image separation or the radius of Einstein ring, in case of perfect alignment, in a specific strong lensing system depends only on angular diameter distances from observer to the lens, lens to source and to the source, provided a reliable model for the mass distribution within the lens is known. A Singular Isothermal Ellipsoid (SIE) model, in which the projected mass distribution is elliptical (\citealt{Ratnatunga+etal+1999}), is often used for the purpose. As most of the lensing galaxies observed are elliptical (early-type), the SIE model is quite reasonable (\citealt{Kochanek+etal+2000}). A simpler Singular Isothermal Sphere (SIS) model, an SIE with zero ellipticity, is also found to be consistent with the observations (\citealt{Cao+etal+2015, Melia}). The Einstein ring radius $\theta_{E}$ in a SIS lens is (\citealt{Schneider}):
\begin{equation}
\theta_{E} = 4\pi \frac{\sigma_{SIS}^{2}}{c^{2}} \frac{D_{ls}}{D_{s}}
\end{equation}
where c is the speed of light, $\sigma_{SIS}$ is the stellar velocity dispersion in the lensing galaxy and $D_{ls}/D_{s}$ is the ratio of the angular diameter distance between source and lens and between source and observer. If one has the values of $\theta_{E}$ from image astrometery and $\sigma_{SIS}$ from spectroscopy, the distance ratio 
$D^{ob}(z_{l},z_{s}) = \frac{D_{ls}}{D_{s}}$ can be estimated. 

\subsection{Distance Sum Rule}
The ratio of angular diameter distances $D_{ls}$ and $D_{s}$ in equation ($6$) can be expressed in terms of the corresponding dimensionless co-moving distances $d(z)$ as:
\begin{equation}
D(z_{l},z_{s}) = \frac{d_{ls}}{d_{s}}
\end{equation}
If $d(z)$ is monotonic and $d'(z) >0$, then $d_{ls}$ and $d_{s}$ are related by the distance sum rule (\citealt{Peebles, Ras}):
\begin{equation}
d_{ls} = d_{s}\sqrt{1+\Omega_{K} d^{2}_{l}} - d_{l}\sqrt{1+\Omega_{K} d^{2}_{s}}
\end{equation}
Therefore,
\begin{equation}
D(z_{l},z_{s}) = \sqrt{1+\Omega_{K} d^{2}_{l}} - \frac{d_{l}}{d_{s}}\sqrt{1+\Omega_{K} d^{2}_{s}}
\end{equation}
Using the distance ratio values  $D(z_{l},z_{s})$, obtained from SGL observations, a continuous distance function $d(z)$ can be constructed in a cosmological model-independent way using a polynomial fit (\citealt{Ras,Collet,Wei2,Qi}). Here, a third order polynomial
\begin{equation}
d(z) = z + a_{1}z^{2} + a_{2}z^{3}
\end{equation} 
with the initial conditions $d(0) = 0$ and $d'(0) =1$, is constructed by fitting the SGL data.

\subsection{SN Ia as Standard Candles}
 Type Ia supernovae, due to their superior brightness, are often adopted to provide the most effective method to measure luminosity distances . The distance modulus $\mu$ of a supernova is defined as:
\begin{equation}
\mu^{th}(z) = m - M' = 5 log(D_{L}/Mpc)+ 25
\end{equation}
where $m$ is the apparent magnitude, $M'$ is the absolute magnitude and $D_{L}$ is the luminosity distance.
Using distance duality equation,
\begin{equation}
\mu^{th}(z) = m - M' = 5 log(\frac{c}{H_{0}} (1+z) \eta(z) d(z)/Mpc)+ 25
\end{equation}

To test the violation of CDDR, I use the following parametrizations of $\eta(z)$:
$$(i) \;\;\;\;\; \eta(z) = 1 + \eta_{0} z$$
 and
 $$(ii)\;\;\;\;\;  \eta(z) = 1 + \frac{\eta_{0} z}{1+z}$$
 
\section{Data and Methodology}
Strong gravitational lensing data used in this study is taken from the catalog of 161 galaxy scale source SGL systems, compiled by  \citet{Chen+etal+2019},  from the LSD, SL2S, SLACS, S4TM, BELLS and BELLS GALLERY surveys.  Only early-type lens galaxies with E or S0 morphologies which do not have any significant substructure or close massive companion are selected for the sample. These two conditions ensured that the lens galaxies in the sample are spherically symmetric. 
For each lens, the source redshift ($z_{s}$), lens redshift  ($z_{l}$) and luminosity averaged central velocity dispersion measured within the aperture $ \sigma_{ap}$ are determined spectroscopically. As the lenses in the sample are from different surveys, $\sigma_{ap}$ is normalised to the velocity dispersions within circular aperture of radius $R_{eff}/2$, where $R_{eff}$ is the half-light radius of the lens. The normalised velocity dispersion $\sigma_{0}$ is (\citealt{Jor,Jor1,Cap}):
\begin{equation}
\sigma_{0} = \sigma_{ap}(\theta_{eff}/(2\theta_{ap}))^{\nu}
\end{equation} 
where $\theta_{eff} = R_{eff}/D_{l}$ and $\nu$ is the correction factor to be fitted from the samples of observations. Following Jorgensen et al. (\citealt{Jor,Jor1}), I take $\nu = -0.04$. Since it has been established that the intermediate-mass early-type elliptical lens galaxies show the best consistency with the SIS lens model (\citealt{Cao1,Koop,Koop1,Treu}), I select only those systems from the catalog whose velocity dispersion is in the range $200 km s^{-1}\leq \sigma_{ap} \leq 300 km s^{-1}$. As the velocity dispersion for a SIS model $\sigma_{SIS}$ may not be same as the central velocity dispersion $\sigma_{0}$, a new parameter $f_{E}$ was introduced by Kochanek (\citealt{Koch}) such that $\sigma_{SIS} = f_{E}\sigma_{0}$. The parameter $f_{E}$ compensates for the contribution of dark matter halos in velocity dispersion, systematic errors in measurement of image separation and any possible effect of background matter over lensing systems. All these factors can affect the image separation by up to $20\%$ which limits $\sqrt{0.8} < f_{E} < \sqrt{1.2}$ (\citealt{Cao,Ofek}). In the present analysis,  $f_{E}$  is taken as a free-parameter and is fitted  along with the other parameters (\citealt{SG}). 

 The Einstein radius $\theta_{E}$ is determined by fitting model mass distributions to generate model lensed images and comparing them to the observed images taken from Hubble Space Telescope or from earth based telescopes. The relative uncertainty in $\theta_{E}$ is taken to be $5\%$ for all lenses (\citealt{Cao1}). Finally, from the  catalog of SGL systems 
I also exclude the systems for which the distance ratio $D^{ob} >1$, as they are  not physical (see equation (8)). The distribution of data points in the catlog with $\sigma_{ap}$ and $D^{ob}$ is shown in Fig.1.

\begin{figure}[p]
\centering

  \includegraphics[width=0.5\linewidth]{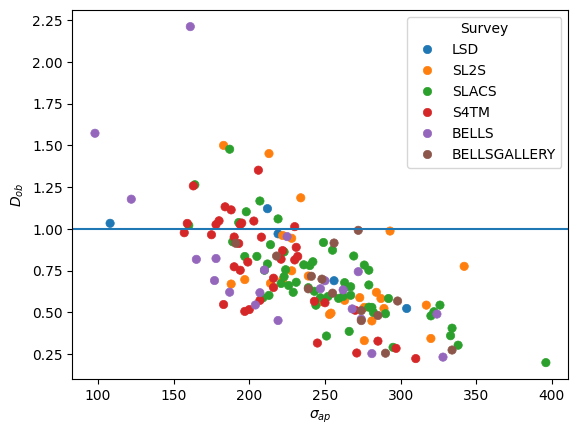}
 \caption{ \small{From the catalog of 161 SGL systems, only those systems which show best consistency with the SIS lens model (for which $200 km s^{-1}\leq \sigma_{ap} \leq 300 km s^{-1} $) are chosen. Further, the lens systems with the observed distance ratio $D_{ob} > 1$ (being unphysical) are excluded. }}
  
\end{figure}   

The likelihood function for the SGL dataset is:
\begin{equation}
L = e^{-\chi_{1}^{2}/2}
\end{equation}
where:
\begin{equation}
\chi_{1}^{2} = \sum_{i=1}^{N} \left(\frac{D^{th}(z_{l,i},z_{s,i},a_{1},a_{2})-D^{ob}(\sigma_{0,i},\theta_{E,i})}{D^{ob}\Delta D^{ob)_{i}}}\right)^{2}
\end{equation}
here $N = 102$ is the number of data points for which  $D^{ob} < 1$ and velocity dispersion is in the range $200 km s^{-1}\leq \sigma_{ap} \leq 300 km s^{-1}$. $\Delta D^{ob}$ is the uncertainty in the value of $D^{ob}$ given by:
\begin{equation}
\Delta D^{ob} = \sqrt{(\frac{\Delta\theta_{E}}{\theta_{E}})^{2}+ 4(\frac{\Delta\sigma_{0}}{\sigma_{0}})^{2}}
\end{equation}
where $\Delta\theta_{E}$ and $\Delta\sigma_{0}$ are the uncertainties  in the measurement of Einstein radius and velocity dispersion respectively. 

The Pantheon+  sample, recently  published by \citet{Scolnic}, is used to estimate the luminosity distance $D^{ob}_{L}(z)$. Scolnic et al. reported the corrected apparent magnitude $m^{corr} = \mu^{ob} + M_{B}$ by analysing 1701 supernova light curves from 1550 distinct type Ia supernovae ranging in redshift from $z = 0.001$ to $2.26$ and fitting them  using a SALT2 model. To reduce the systematic uncertainties of the dataset, BEAMS method with bias corrections is used (\citealt{Kess+Scol+17}) to make corrections to selection biases and contamination from core-collapse SNe.
The uncertainties, including statistical and systematic uncertainties, for the data are given by the $1701$x$1701$ covariance matrix $C$ (\citealt{Brout,Riess}). The covariance matrix $C$ accounts for the errors from the measurement of redshifts, peculiar velocities of host galaxies, calibration of light curves and the SALT2 model fitting, extinction due to the Milky Way, and simulations of survey modelling, distance modulus uncertainty modelling, and intrinsic scatter models.

The chi-square function for the supernovae Pantheon+ sample is:
\begin{equation}
\chi_{2}^{2} =  \mathbf{{\bigtriangleup\mu}^{\dagger} \cdot C^{-1} \cdot \mathbf{\bigtriangleup\mu}}
\end{equation}
where $\bigtriangleup\mu_{i} = \mu^{th}(z_{i},\textbf{p}) -  \mu^{ob}(z_{i})$ is the vector of residuals of the sample and $\textbf{p} = (\Omega_{K},\eta_{0},a_{1},a_{2},M)$ is the vector formed by the parameters to be fitted from the data. Given the degeneracy of factor $5 log_{10}(c/ H_{0}Mpc)+25$ with absolute magnitude $M'$, both are combined and fitted as the parameter $M =  M' +5 log_{10}(c/ H_{0}Mpc)+25$. 

The combined log-likelihood function for SGL and SNe datasets is:
\begin{equation}
\textit{ln}(L_{tot}) = -0.5(\chi_{1}^{2} + \chi_{2}^{2})
\end{equation}
The best fitting values of the parameters can be obtained by maximizing the likelihood function $L_{tot}$.

\section{Results and Conclusions}
I sample the likelihood function $L_{tot}$  by performing a Markov Chain Monte Carlo (MCMC) analysis using Python module \textit{emcee} (\citealt{emcee}) and compute the 1-dimensional marginalized best fitting values and $68\%$ uncertainties of the parameters. The results are listed in Table 1. The 2-dimensional contours and the 1-dimensional posterior probability distributions for the parameters, generated using the Python module GETDIST (\citealt{getdist}), are shown in Fig. 2-5.

$\;\;\bullet$ In the spatially flat Universe ($\Omega_{k} = 0$), with linear parametrization i. e. $\eta(z) = 1+ \eta_{0} z$, the 1-D marginalized best fit value of $\eta_{0}$ is $-0.0051_{-0.0621}^{+0.0677}$ which is in excellent agreement  with the CDDR at a very high level of confidence.
With non-linear parametrization for $\eta(z)$, the 1-D marginalized best fit value of $\eta_{0}$ is $-0.0921_{-0.0663}^{+0.0778}$. The value of $\eta_{0}$ shows no significant deviation from CDDR.

$\;\;\bullet$ With a non-flat geometry, the 1-D marginalized best fit value of $\eta_{0}$ for linear and non-linear parametrizations are: 
 $0.033_{-0.0702}^{+0.0749}$ and  $-0.1205_{-0.0740}^{+0.0809}$ respectively.
Again the value of $\eta_{0}$ with linear parametrization is in agreement  with the CDDR within $1\sigma$ confidence interval. With non-linear parametrization too  CDDR is true within 2$\sigma$ confidence interval. The data set provides a weak constraint on $\Omega_{K}$ but CDDR holds for non-flat Universe with tight constraints on $\eta_{0}$ which is  mildly dependent on $\Omega_{K}$.

 For a comparison, the results of some of the works with a model independent approach are presented in Table 2.
\citet{Ruan+etal+2018} tested CDDR based on strong gravitational lensing and a reconstruction  of the $H II$ galaxy Hubble diagram  using Gaussian processes.
\citet{Xu} used  the combination of BAO measurements and SNe Ia sample. They applied both artificial neural network (ANN) method and binning the SNe Ia sample method to derive the values of luminosity distance at the redshifts of BAO measurements. Whereas,  \citet{Wang} derived luminosity distance values at the redshifts of BAO
measurements by binning and using  Gaussian processes.
\citet{Lyu+etal+2020} and \citet{Liao+etal+2016}  tested CDDR by combining the SGL  observations with SNe Ia data using linear parametrization for $\eta(z)$. They combine each SGL system with two SNe Ia events closest to the redshift of the source and lens with a difference not exceeding $0.005$. Using the SIS lens mass model, \citet{Lyu+etal+2020} reported $\eta_{0} = -0.161^{+0.062}_{-0.058}$, showing a moderate tension with the CDDR. 
 In contrast, by choosing only those SGL systems which are found to be in better accordance with the SIS lens model and using the larger and updated SNe Ia data (Pantheon+), $\eta_{0 }$ is consistent with $0$ with high precision. The issue of finding the supernovae from the Pantheon+ dataset whose redshift match with that of lens and source in each SGL system is overcome by reconstructing a continuous co-moving distance function $d(z)$,  that best approximates the discrete observed data, using a polynomial fit. The best fit value of $\eta_{0}$ in spatially flat Universe is consistent with the values reported in other works and with similar or better precision (Table 2).
 In addition, the method also confirms the validity of CDDR with a small confidence interval in the curved space-time.

\begin{table}

With $\eta(z) = 1 + \eta_{0} z$

\begin{tabular}{|c|c|c|c|c|c|}  \hline\hline\

$\Omega_{K}$ & $\eta_{0}$ & $a_{1}$ & $a_{2}$ & $f_{E}$ & M  \\ 
\hline & & & & & \\
0&$-0.0051_{-0.0621}^{+0.0677}$ & $-0.2942^{+0.0641}_{-0.0664}$ & $0.0471^{+0.0176}_{-0.0154}$ &$1.0033^{+0.0109}_{-0.0109}$&$23.8226^{+0.0063}_{-0.0066}$ \\ 
&&&&&\\
$0.0567^{+0.0313}_{0.0374}$ & $-0.0330^{+0.0749}_{-0.0702}$ & $-0.2681^{+0.0712}_{-0.0731}$ & $0.0438^{+0.0174}_{-0.0149}$ & $ 1.0001^{+0.0118}_{-0.0106} $ & $23.8233^{+0.0059}_{-0.0068}$\\
&&&&&\\

&&&&&\\
 \hline \hline
\end{tabular}

With $\eta(z) = 1 + \eta_{0} (z/1+z)$:

\begin{tabular}{|c|c|c|c|c|c|}  \hline\hline\
$\Omega_{K}$ & $\eta_{0}$ & $a_{1}$ & $a_{2}$ & $f_{E}$ & M  \\ 
\hline & & & & & \\
0&$-0.0921_{-0.0663}^{+0.0778}$ & $-0.2428^{+0.0421}_{-0.0486}$ & $0.0301^{+0.0162}_{-0.0132}$ &$0.9984^{+0.0083}_{-0.0082}$&$23.8286^{+0.0088}_{-0.0081}$ \\ 
&&&&&\\
$0.1485^{+0.0948}_{-0.0952}$ & $-0.1205^{+0.0809}_{-0.0740}$ & $-0.2292^{+0.0519}_{-0.0500}$ & $0.0278^{+0.0167}_{-0.0162}$ & $ 1.0017^{+0.0086}_{-0.0092} $ & $23.8309^{+0.0086}_{-0.0089}$\\
&&&&&\\

&&&&&\\
 \hline \hline
\end{tabular}

\caption{\small 1-D marginalized best fit parameter values and uncertainties.}

\end{table}

\begin{table}

\begin{center}

\begin{tabular}{|c|c|c|c|}  \hline\hline\
 Data & $\eta_{0}(L)$ & $\eta_{0}$ (N-L) & Ref. \\ 
\hline & & & \\

SNe Ia + SGL & $-0.005^{+0.351}_{-0.251}$ &-&\citet{Liao+etal+2016} \\
H II + SGL & $0.0147^{+0.056}_{-0.066}$ &-& \citet{Ruan+etal+2018} \\
SNe Ia + SGL & $-0.161^{+0.062}_{-0.058}$ &-& \citet{Lyu+etal+2020} \\
BAO + SNe Ia  & $-0.064^{+0.057}_{-0.052}$ & $-0.181^{+0.160}_{-0.141}$&\citet{Xu} \\
BAO + SNe Ia  & $0.041^{+0.123}_{-0.109}$ & $0.082^{+0.246}_{-0.214}$ &\citet{Wang} \\
SNe Ia + SGL  & $-0.0051^{+0.0677}_{-0.0621}$ & $-0.0921^{+0.0778}_{-0.0663}$ & This work\\
&&&\\
\hline \hline
\end{tabular}

\caption{\small Best fit values of $\eta_{0}$ from other independent studies.}
\end{center}
\end{table}

\begin{figure}[p]
\centering

  \includegraphics[width=1.0\linewidth]{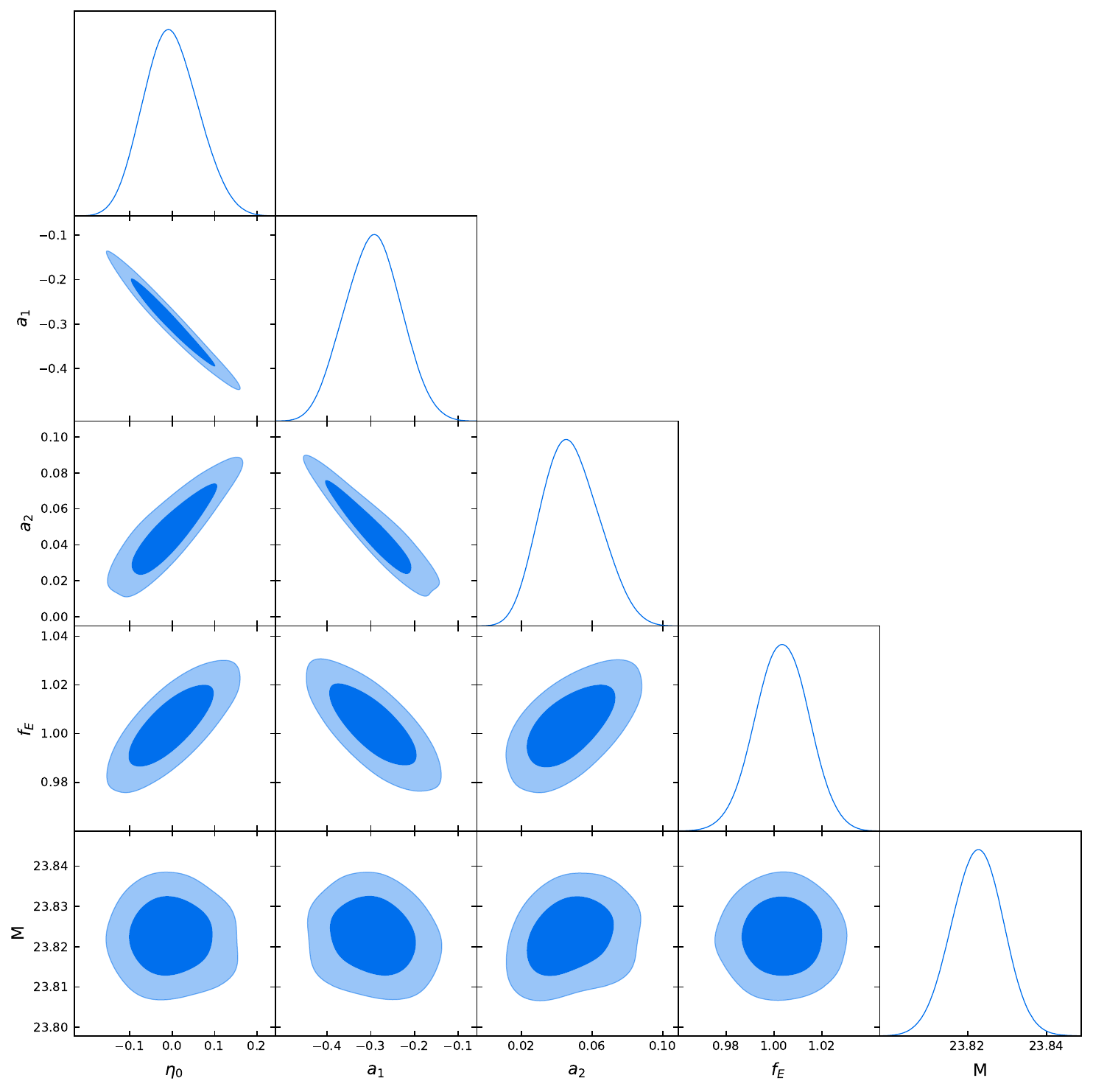}
 \caption{ \small{ 1-D posterior probability distributions and 2- D confidence  regions of the parameters in flat spacetime with $\eta(z) = 1 + \eta_{0} z$.}}
  
\end{figure}
\begin{figure}[p]
\centering

  \includegraphics[width=1.0\linewidth]{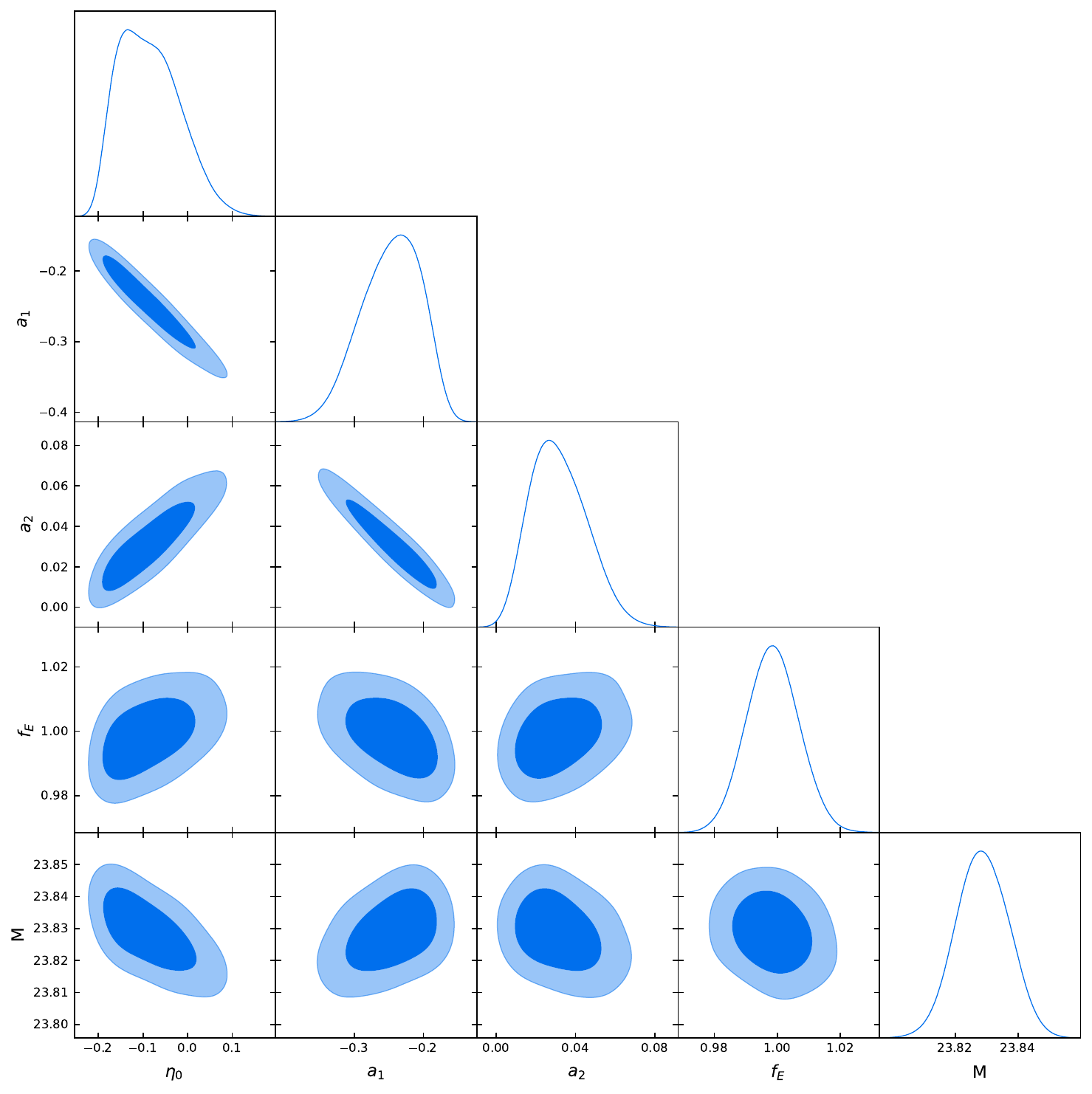}
 \caption{ \small{ 1-D posterior probability distributions and 2- D confidence  regions of the parameters in flat spacetime with $\eta(z) = 1 + \eta_{0} (z/1+z)$.}}  
\end{figure}

\begin{figure}[p]
\centering

  \includegraphics[width=1.0\linewidth]{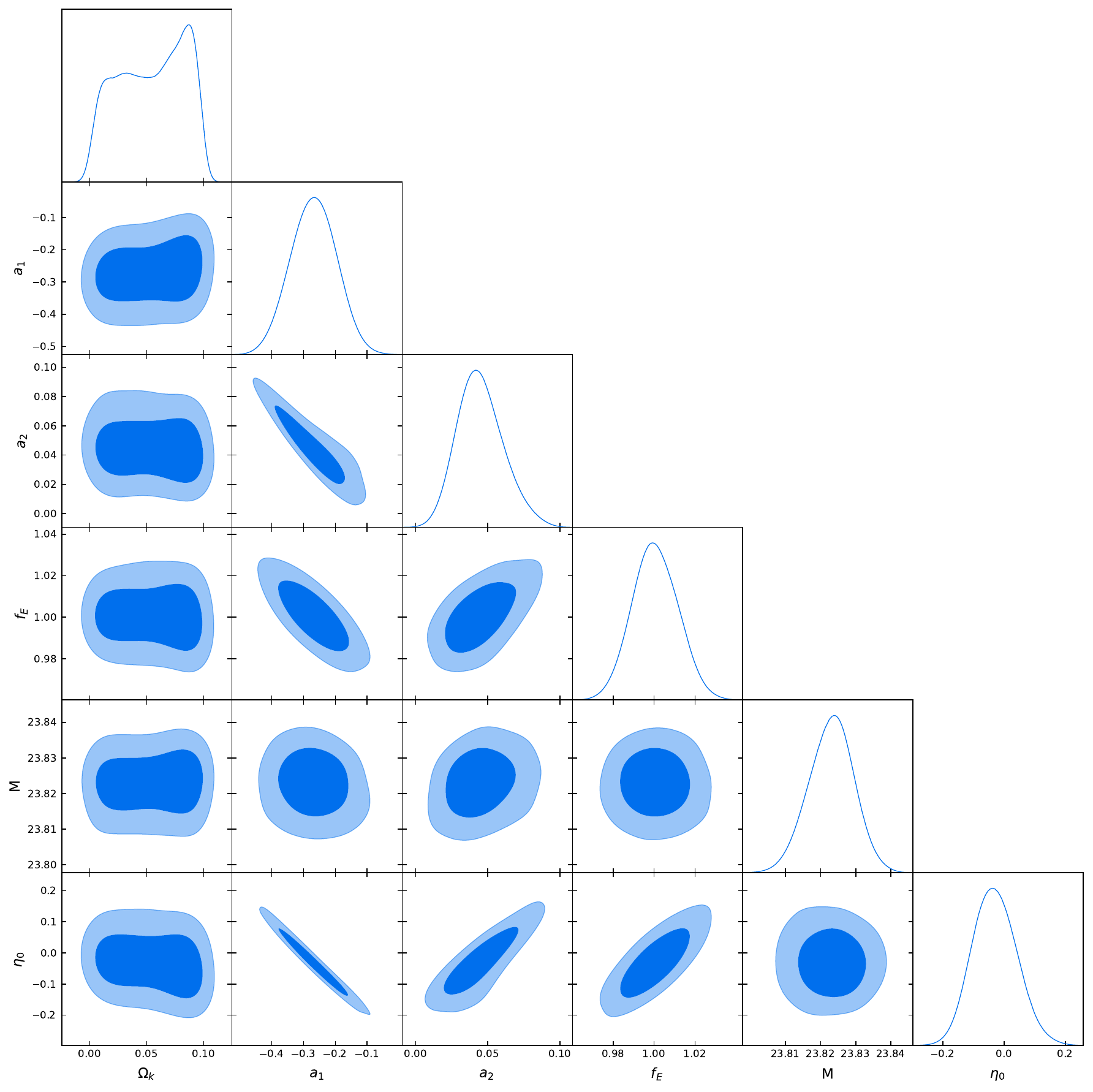}
 \caption{ \small{ 1-D posterior probability distributions and 2- D confidence  regions of the parameters in non- flat spacetime with $\eta(z) = 1 + \eta_{0} z$.}}  
\end{figure}

\begin{figure}[p]
\centering

  \includegraphics[width=1.0\linewidth]{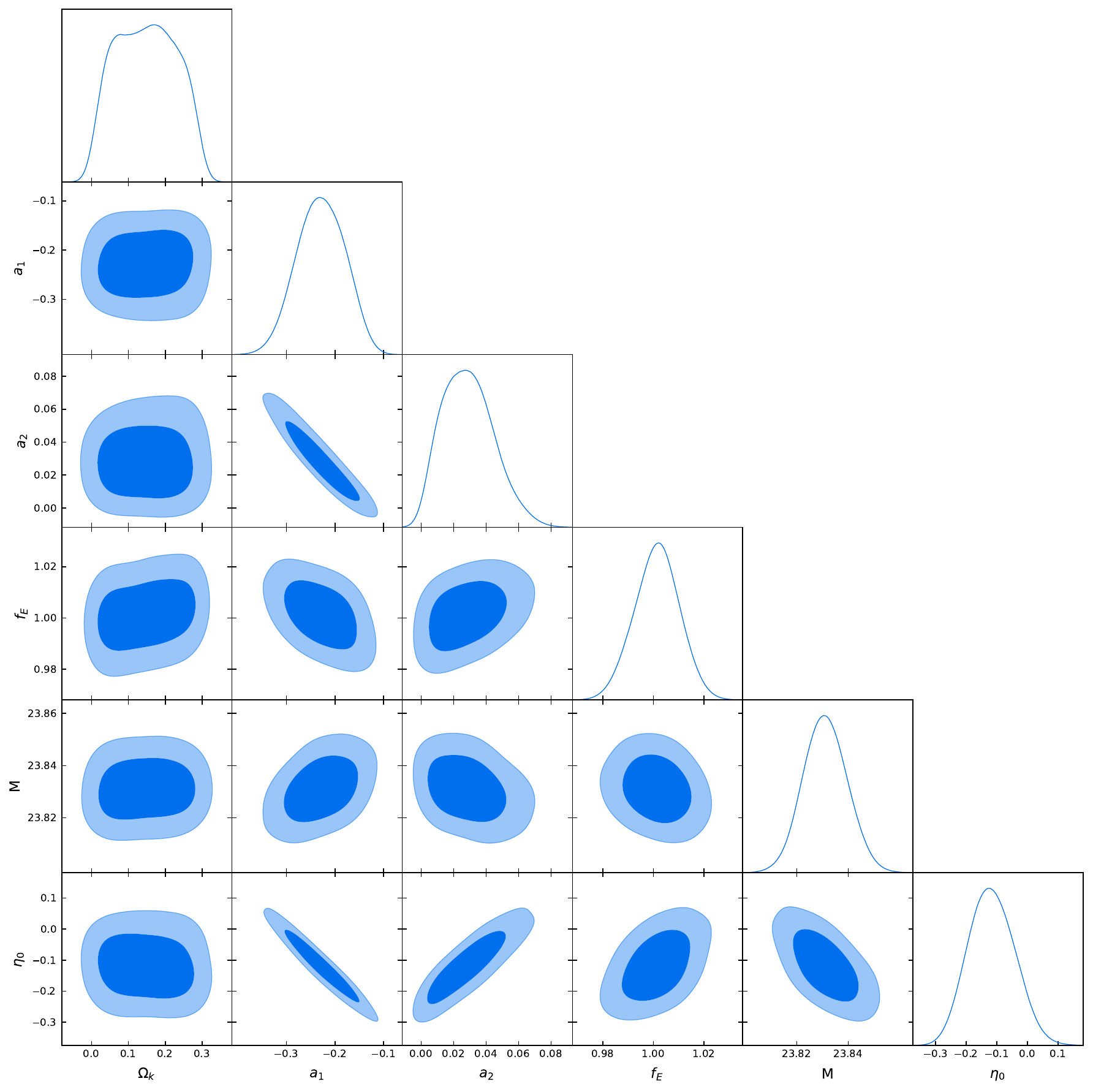}
 \caption{ \small{ 1-D posterior probability distributions and 2- D confidence  regions of the parameters in non-flat spacetime with $\eta(z) = 1 + \eta_{0}(z/1+z) $.}}  
\end{figure}

\section{Discussion}
Cosmic distance duality relation is one of the fundamental relations in cosmology which  plays a crucial role in astronomical observations. Any deviation from CDDR indicates that either the spacetime is not described by a metric theory of gravity or new  physics beyond the one we understand. With  the  availability of new improved observational data with better precision, it is important to test this relation. In this paper, using the latest SGL and SNe Ia Pantheon+ data jointly, I probe the validity of CDDR without considering any cosmological model. In most of the studies of CDDR so far, spatial  flatness of space-time is assumed. Whereas,  using the distance sum rule along null geodesics in the FLRW metric, I test CDDR in a non-flat space-time as well as flat space-time. The analysis shows that CDDR is valid at a very high level of confidence in the flat  space-time. In a non-flat space-time too, CDDR is true within $1\sigma$ confidence interval with a mild dependence of $\eta_{0}$ on $\Omega_{K}$. To check the effect of parametrization, a non-linear parametrization for $\eta (z)$ is used and found no significant deviation from CDDR.  
The high level of confidence achieved in the results is attributed to the improved SNe Ia observations with the covariance matrix incorporating all kinds of statistical and systematic uncertainties. Further,  new powerful space and ground-based telescopes for imaging and spectroscopic observations helped to achieve better precision for SGL observations. The only limitation while using SGL data is the uncertainty  about the mass distribution in the lens. To mitigate the impact of an imprecisely known mass distribution, I selected only those lensing systems which are proved to be in accordance with the SIS lens model.  Also, the model independent method used in the work helped to avoid the unknown systematics associated with the models.

The results for both flat and non-flat space-time establishes the CDDR all the way out to $z\sim 2.3$. However, it is important to test CDDR at higher redshifts up to the last scattering surface. Recently a large number of high redshift quasars are detected with redshifts between 4.44 and 6.53, using the Dark Energy Spectroscopic Instrument (DESI) \citep{yang23}. The multiple measurements of high-redshift quasars can be used to estimate luminosity distances from the relation between the UV and X-ray luminosities of quasars. Whereas, angular diameter distances can be obtained from the compact structures in intermediate luminosity radio quasars. Thus, two different cosmological distances from the same kind of objects at high redshifts can be determined and compared \citep{zheng20}. Further, recent detection of gravitational wave (GW) signals by LIGO and Virgo detectors has heralded the start of GW astronomy and multi-message astronomy era \citep{abbot16a, abbot16b, abbot17}. These “standard sirens” could provide the absolute luminosity distance of the event without any calibration, as GWs propagate freely through a perfect fluid without any absorption or dissipation \citep{Schutz86}. Combining the GW data with SGL or BAO data (for angular diameter distances) can give a more precise test of CDDR.
Theoretically, third generation  GW detectors, Einstein  Telescope (ET), could detect GW signals up to redshift $z \sim 2$ for the neutron star - neutron star mergers and $z \sim 5$ for black hole - neutron star mergers systems \citep{Cai}. Also, a possible detection of strong gravitational wave lensing can  provide simultaneous measurements of both the luminosity and angular diameter distances  which in turn can be used to probe CDDR \citep{Arjona}. However, we should wait for the data from ET until the statistics and the redshift coverage will be sufficient to get competitive results.

I thank Kshitiz Singh for helping me with the codes and 
the anonymous referee for constructive comments to improve the manuscript.

\section*{Declarations}

\textbf{Data Availability Statement:} The datasets used in the work are available in public domain. For SGL data, check https://doi.org/10.1093/mnras/stz1902.
Pantheon+ SNe Ia sample is available at:

https://github.com/PantheonPlusSH0ES/PantheonPlusSH0ES.github.io.\\
\textbf{Funding and/or Conflicts of interests/Competing interests} This work is done independently and no funds, grants, or other support was received.  There are no conflicts of interests that are relevant to the content of this article.

\label{lastpage}


\begin{thebibliography}{99}
\bibitem[Abbot et al. (2016a)]{abbot16a} Abbot B. P., et al., 2016, Phys. Rev. Lett., 116, 061102.

\bibitem[Abbot et al. (2016b)]{abbot16b} Abbot B. P., et al., 2016, Phys. Rev. Lett., 116, 241103.

\bibitem[Abbot et al. (2017)]{abbot17} Abbot B. P., et al., 2017, Phys. Rev. Lett., 118, 221101.

\bibitem[Arjona et al. (2021)]{Arjona} Arjona R., et al., 2021, Phys. Rev. D 103, 103513. 

\bibitem[Basset $\&$ Kunz (2004)]{Basset} Bassett B. A. $\&$  Kunz M., 2004, Phys. Rev. D 69, 101305.

\bibitem[Brout et al. (2022)]{Brout}  Brout D., et al., 2022, ApJ, 938,110.

\bibitem[Cai $\&$ Yang (2017)]{Cai} Cai R. G. $\&$ Yang T., 2017, Phys. Rev. D 95, 044024.

\bibitem[Cao et al. (2012)]{Cao}  Cao S. , et al., 2012, JCAP, 3, 16.

\bibitem[Cao et al. (2015)]{Cao+etal+2015}  Cao S., et al., 2015, ApJ, 806,185.

\bibitem[Cao et al. (2016a)]{Cao1}  Cao S.,  Biesiada M.,  Yao M., $\&$  Zhu Z.-H., 2016, MNRAS, 461, 2192.

\bibitem[Cao $\&$ Zhu (2011)]{Cao11} Cao S. $\&$ Zhu Z. H., 2011,  Sci. China Phys. Mech. Astron. 54, 2260–2264.

\bibitem[Cao et al. (2016b)]{Cao16} Cao S.,  Biesiada M.,  Zheng X., $\&$  Zhu Z.-H., 2016, MNRAS, 457, 281.

\bibitem[Cappellari et al. (2006)]{Cap}  Cappellari M.,  Bacon R.,  Bureau M., et. al. , 2006, MNRAS, 366, 1126.

\bibitem[Chen et al. (2019)]{Chen+etal+2019}  Chen Y.,  Li R.$\&$  Shu Y., 2019, MNRAS, 488, 3745.

\bibitem[Collet et al. (2019)]{Collet} Collet T.,  Montanari F. $\&$  R$\ddot{a}$s$\ddot{a}$nen S., 2019, Phys. Rev. Lett., 123, 231101.

\bibitem[Corasaniti (2006)]{Corasaniti} Corasaniti P. S., 2006, MNRAS, 372, 191. 

\bibitem[Ellis et al. (2013)]{Ellis+etal+2013}  Ellis G. F. R., et al., 2013,  Phys. Rev. D 87, 103530.

\bibitem[Etherington (1933)]{Etherington+1933}  Etherington I., 1933,  The London Edinburgh, and Dublin Philosophical Magazine and Journal of Science, 15, 761.

\bibitem[Foreman-Mackey et al. (2013)]{emcee}  Foreman-Mackey D.,  Hogg D. W.,  Lang D. ,  Goodman J., 2013, PASP, 125, 306.

\bibitem[Gahlaut (2024)]{SG} S. Gahlaut S., 2024, Pramana - J. Phys., 98, 22.

\bibitem[Holanda $\&$ Busti (2014)]{Holanda+Busti+2014}  Holanda R. F. L. $\&$  Busti V. C., 2014, Phys. Rev. D 89, 103517.

\bibitem[Holanda et al. (2011)]{Hol11} Holanda R. F. L., Lima J. A. S. $\&$ Ribeiro M. B., 2011, Astron. $\&$ Astrophys., 528, L14.

\bibitem[Holanda et al. (2012)]{Hol12} Holanda R. F. L., Lima J. A. S. $\&$ Ribeiro M. B. 2012, Astron. Astrophys. 538, A131.

\bibitem[Holanda et al. (2013)]{Holanda} Holanda R. F. L., Carvalho J. C. $\&$  Alcaniz J. S., 2013, JCAP, 04, 027.

\bibitem[Jorgensen et al. (1995a)]{Jor} Jorgensen I.,  Franx M., $\&$  Kjaergaard P., 1995, MNRAS, 273, 1097.

\bibitem[Jorgensen et al. (1995b)]{Jor1} Jorgensen I.,  Franx M., $\&$  Kjaergaard P., 1995 MNRAS, 276, 1341.

\bibitem[Kessler $\&$ Scolnic (2017)]{Kess+Scol+17} Kessler R. $\&$ Scolnic D., 2017, ApJ, 836, 56.

\bibitem[Kochanek (1992)]{Koch} Kochanek C. S., 1992, ApJ 397,381.

\bibitem[Kochanek et al. (2000)]{Kochanek+etal+2000}  Kochanek C. S., et al., 2000, ApJ, 543, 131.

\bibitem[Koopmans et al. (2006)]{Koop}  Koopmans L. V. E., et al., 2006,  ApJ, 649, 599.

\bibitem[Koopmans et al. (2009)]{Koop1}  Koopmans L. V. E., et al., 2006, ApJ, 703, L51.

\bibitem[Lewis (2019)]{getdist}  Lewis A., 2019, "GetDist: a Python package for analysing Monte Carlo samples", arXive e-prints [1910.13970].

\bibitem[Li et al.(2011)]{Li+etal+2011}  Li Z.,  Wu P.$\&$  Yu H., 2011, ApJL 729, L14.

\bibitem[Li et al.(2013)]{Li+etal+2013}  Li Z., et al., 2013, Phys. Rev. D 87, 103013.

\bibitem[Liao et al.(2016)]{Liao+etal+2016}  Liao K., et al., 2016,  ApJ 822, 74.

\bibitem[Lyu et al (2020)]{Lyu+etal+2020}  Lyu M. Z.,  Li Z. X.$\&$  Xia J.-Q., 2020, ApJ, 888, 32.

\bibitem[Melia (2018)]{Melia18} Melia F., 2018, arXive:1804.09906.

\bibitem[Melia et al. (2015)]{Melia}  Melia F.,  Wei J.-J. $\&$  Wu X.-F., 2015, AJ, 149, 2.

\bibitem[Nesseris $\&$ Bellido (2012)]{Nesseris+Bellido+2012}  Nesseris S. $\&$  Bellido J. G., 2012, JCAP 11, 033.

\bibitem[Ofek et al. (2003)]{Ofek} Ofek E. O.,  Rix H. W. $\&$  Maoz D., 2003, MNRAS 343, 639.

\bibitem[Peebles (1993)]{Peebles}  Peebles P, J. E., 1993, "Principals of Physical Cosmology", Princeton University Press.

\bibitem[Qi et al. (2021)]{Qi}  Qi J.- Z., et al., 2021, MNRAS, 503, 2179.

\bibitem[Rana et al. (2016)]{Rana+etal+2016}  Rana A., et al., 2016,  JCAP ,07 , 026.

\bibitem[Rana et al (2017)]{Rana+etal+2017} Rana A. et. al., 2017, JCAP, 010.

\bibitem[R$\ddot{a}$s$\ddot{a}$nen et al. (2015)]{Ras} R$\ddot{a}$s$\ddot{a}$nen S.,  Bolejko K.$\&$  Finoguenov A., 2015, Phys. Rev. Lett., 115,101301.

\bibitem[Ratnatunga et al., (1999)]{Ratnatunga+etal+1999}  Ratnatunga K. U.,  Griffiths R. E. $\&$  Ostrander E. J., 1999, AJ, 117, 2010.

\bibitem[Riess et al. (2022)]{Riess}  Riess A. G., et al., 2022,  ApJ Lett., 934, L7.

\bibitem[Ruan et al (2018)]{Ruan+etal+2018}  Ruan C.-Z.,  Melia F. $\&$  Zhang T.-J., 2018, ApJ, 866, 31.

\bibitem[Schneider et al. (2006)]{Schneider}  Schneider P.,  Kochanek C., $\&$  Wambsganss J., 2006, :"Gravitational Lensing:
Strong, Weak and Micro", Springer-Verlag Berlin Heidelberg, eBook-ISBN 978-3-540-30310-7.

\bibitem[Schutz (1986)]{Schutz86} Schutz B. F., 1986, Nature, 323, 310.

\bibitem[Scolnic et al. (2022) ]{Scolnic}  Scolnic D., et al., 2022, ApJ, 938, 113 6.

\bibitem[Shafieloo et al. (2013)]{Shafieloo+etal+2013}  Shafieloo A.,  Majumdar S.,  Sahni V. $\&$  Starobinsky A. A., 2013, JCAO, 04 , 042.

\bibitem[Treu et al. (2006)]{Treu}  Treu T., et al.,2006,  ApJ, 650, 1219.

\bibitem[Wang et al. (2024)]{Wang}  Wang M.,  Fu X.,  Xu B., et al., 2024, Eur. Phys. J. C 84, 702.

\bibitem[Wei $\&$ Melia (2020)]{Wei2}  Wei J.- J. $\&$  Melia F., 2020, ApJ 897, 127.

\bibitem[Xu et al. (2022)]{Xu}  Xu B.,  Wang Z.,  Zhang K., et al., 2022, ApJ, 939, 115.

\bibitem[Yang et al. (2023)]{yang23} Yang J., et al., 2023, ApJS, 269, 27.

\bibitem[Zheng et al. (2020)]{zheng20} Zheng X., et al., 2020 ApJ 892, 103.








\end{thebibliography}
\end{document}